%% file: main.tex
\documentclass[12pt, draftcls, onecolumn]{IEEEtran}
\usepackage{graphicx} % for pdf, bitmapped graphics files
\usepackage{epsfig} % for postscript graphics files
\usepackage{mathptmx} % assumes new font selection scheme installed
\usepackage{times} % assumes new font selection scheme installed
\usepackage{amsmath} % assumes amsmath package installed
\usepackage{amssymb}  % assumes amsmath package installed
\usepackage{color}
\usepackage{authblk}
\usepackage[caption=false,font=footnotesize]{subfig}
\usepackage{setspace}
\usepackage{algorithm}
\usepackage{algorithmic}
\usepackage[top=0.75in, bottom=0.6in, left=0.6in, right=0.5in]{geometry} 
\def\proof{\noindent{{\bf Proof: }}}

\newtheorem{theorem}{Theorem}
\newtheorem{lemma}{Lemma}

\newcommand{\E}[1]{{\mathbb{E}}\left[{#1}\right]}

\renewcommand\boldsymbol[1]{\pmb{#1}}

\begin{document}
%\onehalfspacing
\title{{Motivating Workers in Federated Learning: A Stackelberg Game Perspective}}%{\thanks{This material is based upon work supported by the National
%Science Foundation under Grants CNS-0831919, CCF-0916664, CAREER-1054738.}}\thanks{Portions of this work were presented at Asilomar Conference on Signals, Systems, and Computers (Asilomar '10), Pacific Grove, CA.}}
%\author{\IEEEauthorblockN{C.~Emre Koksal}
%\IEEEauthorblockA{Department of Electrical and Computer Engineering,\\
%The Ohio State University\\
%Columbus, OH \\
%Email: koksal@ece.osu.edu} \and \IEEEauthorblockN{Ozgur Ercetin and
%Yunus Sarikaya}
%\IEEEauthorblockA{Faculty of Engineering and Natural Sciences,\\
%Sabanci University,\\
%Istanbul, TR.\\
%Email: \{oercetin,ysarikaya\}@sabanciuniv.edu}}

\author{{Yunus Sarikaya and Ozgur Ercetin}%
\thanks{Y. Sarikaya (ysarikaya@sabanciuniv.edu) and O. Ercetin (oercetin@sabanciuniv.edu)  are with the Department of Electronics Engineering, Sabanci University, Turkey.} }%
%\thanks{C.~E. Koksal (koksal@ece.osu.edu) is with the Department of Electrical and Computer Engineering at The Ohio State University, Columbus, OH.}  \thanks{O. Ercetin (email: oercetin@sabanciuniv.edu) and  Y.Sarikaya (email: sarikaya@su.sabanciuniv.edu) are with the Department of Electronics Engineering, Faculty of Engineering and Natural Sciences, Sabanci University, 34956 Istanbul, Turkey.}}

\maketitle

\input{abstract}
\vspace{-0.1in}
{\allowdisplaybreaks\input{intro}}
\vspace{-0.1in}
{\allowdisplaybreaks\input{system_model}}
%{\allowdisplaybreaks\input{control}}
%{\allowdisplaybreaks\input{section_general}}
%\vspace{-0.1in}
{\allowdisplaybreaks\input{numerical}}
%\vspace{-0.1in}
\input{conclusion}
%\tiny
%\vspace{-0.1in}
\vspace{-0.1in}
\bibliographystyle{IEEEtran}
\bibliography{Learning,macros_abbrev}

\appendices
%\vspace{-0.05in}
\vspace{-0.1in}
{\allowdisplaybreaks\input{appendix_boundary}}
\vspace{-0.1in}
{\allowdisplaybreaks\input{appendix_Theorem}}
%\vspace{-0.1in}

%\input{appendix_recursive}

\end{document}

%% file: abstract.tex
\begin{abstract}
 %Due to the privacy concerns as well as the large size of the training data, distributed learning approaches such as federated learning have gained attention recently. However, the convergence rate of  distributed learning suffers from heterogeneous worker performance, i.e., the model owner has to wait for the slowest machine in each synchronous batch. In this paper, we consider an incentive mechanism for workers to mitigate the delays in completion of each batch.  We analytically obtained Nash equilibrium solution of a Stackelberg game where the model owner is the leader and users are followers.  We numerically evaluated the performance of this game over MNIST dataset. Our results indicate that with a limited budget, the model owner should judiciously decide on the number of workers since there is a trade off between the diversity provided by the number of workers and the latency of completing the training. 
 
  Due to the large size of the training data, distributed learning approaches such as federated learning have gained attention recently. However, the convergence rate of distributed learning suffers from heterogeneous worker performance. In this paper, we consider an incentive mechanism for workers to mitigate the delays in completion of each batch.  We analytically obtained equilibrium solution of a Stackelberg game. Our numerical results indicate that with a limited budget, the model owner should judiciously decide on the number of workers due to trade off between the diversity provided by the number of workers and the latency of completing the training. 
\end{abstract}

%% file: intro.tex
\section{Introduction}
 The recent success of deep learning approaches for domains such as speech recognition  and computer vision stems from many algorithmic improvements but also from the fact that the size of available training data has grown significantly over the years, together with the computing power. The current trend is to use a larger data set and to train deeper networks (higher number of layers) to improve the accuracy. However, the complexity and the memory requirements quickly become unmanageable within the resources of a single machine. An efficient way to deal with this colossal computing task within a reasonable training time is to adopt distributed computation, and to exploit computation and memory resources of multiple machines in parallel.  In \cite{McMahan16}, the federated learning system was introduced to allow the mobile devices perform computation of model training locally on their training data according to the model released by the model owner. Such a design enables mobile users to collaboratively learn a shared prediction model while keeping all the training data private on the device. 
 
 Most of the popular distributed training algorithms include mini-batch versions of stochastic gradient descent (SGD).
 %and other stochastic optimization algorithms such as AdaGrad, RMSProp, and ADAM. 
 Unfortunately, bulk- synchronous implementations of stochastic optimization are often slow in practice due to the need to wait for the slowest machine in each synchronous batch, i.e., they suffer from the so called {\em straggler effect}. For example, experiments on Amazon EC2 instances show that some workers can be five times slower than the typical performance \cite{amazon}.  There have been several attempts in the literature to mitigate the straggler effect by adding redundancy to the distributed computing system via coding \cite{Lee:TIT:18,Karakus17} or via scheduling computation tasks \cite{Harlap16, Sun:arxiv:19}.  
 %Separately, existing works mainly consider a single {\em master} model owner. However, in practice, workers may be shared by more than one {\em masters} to carry out multiple large-scale computation tasks in parallel, or individual workers may have their own tasks to run.  
 %This case is especially relevant for the fifth generation (5G) of the cellular network architecture, where it is expected that a signiﬁcantly higher number of connected wireless devices will collect huge amount of data and process it closer to the edge to improve the end-to-end latency and to avoid the bandwidth limitations.  
 %
 %The workers in a federated learning system are expected to be independent (and possibly mobile) user devices simultaneously running multiple applications.  
% The proposed approach jointly optimized the number of local
%learning and global update cycles at the learners and the
%orchestrator, respectively. 
However, these works overlooked the inherent heterogeneity in the computing capacity of different workers. It is crucial to consider the implications of such heterogeneity on optimizing the task allocation to different workers, improving learning accuracy, minimizing latency, and/or minimizing energy consumption. In that sense, \cite{Mohammad:arxiv:18} considers the problem of adaptive task allocation with the aim to maximize the
learning accuracy, while satisfying a delay constraint resulting from data
distribution/aggregation over heterogeneous channels, and local
computation on heterogeneous nodes.  Furthermore, the limited computing resource of a user device is shared among all running applications. The independent and rational mobile clients need an incentive to participate in federated learning. Hence, a critical question that needs to be addressed by each worker is ``How much Central Processor Unit (CPU) resource of heterogeneous workers should be allocated to the training task of the model owner?"  The answer to this question has repercussions for the central model owner, since in its most basic form of synchronous SGD, the model owner has to wait for the gradients from all workers in order to update its current set of model parameters.  
 
In \cite{Feng:arxiv:18}, a game theoretical approach is established to consider a communication incentive in federated learning, where the aim was to construct a relay network and cooperative communications for supporting model update transfer. Unlike the previous works, in this paper we consider an incentive-based approach to motivate the workers to allocate more computation power for local training. In this setting, at each gradient update step, the model owner offers an incentive to each worker participating in the federated learning process.  Based on this incentive, the workers determine the CPU power they will use to calculate their gradient from the local data. The model owner has a finite budget and distributes its budget among its workers to achieve a fast convergence to a target error rate. We model the interaction between the mobile devices and the model owner as a Stackelberg game. In Stackelberg game, the model owner is the buyer as it buys the learning service provided by the mobile devices. Then, the mobile devices that are the service providers act as the sellers. The model owner inherently acts as the single leader in the upper level of the Stackelberg game while the mobile devices are the corresponding followers.  We obtained the equilibrium solution of this game by first quantifying the average time required to finish a single iteration of SGD.  We also implemented our game theoretical algorithm numerically on MNIST dataset. Our analysis provides insights on the optimal number of workers to achieve a desired balance of error-latency tradeoff. 
 

%% file: system_model.tex
\section{System Model}
\label{sec:model}

We consider a cooperative federated learning system as shown in Figure 1. Specifically, a model owner employs a set of mobile devices, i.e., workers to train a high-quality centralized model. The workers fetch the current parameters $\boldsymbol{w}_i$ from the model owner as and when instructed in the algorithm. Then, they compute gradients using one mini-batch and push their gradients back to the model owner. At each iteration, the model owner aggregates the gradients computed by the workers and updates the parameter  $\boldsymbol{w}$. 
\begin{figure}[t]
\begin{center} 
\includegraphics[width=4in]{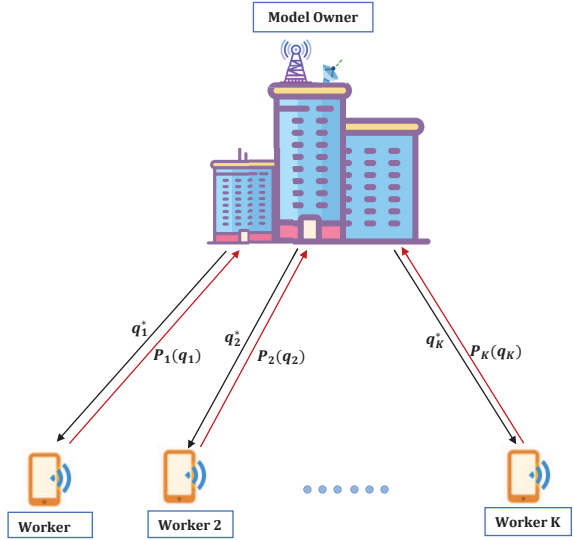}
\label{fig:game_network}
\caption{Game Model of Learning Network}
\end{center}
\end{figure}

Let $T_{i,t}$ be the time elapsed for the worker $i$ to update the gradient in iteration $t$. Here, we consider plain synchronous SGD such that the model owner waits for all the workers to push their gradients. Thus, iteration $t$ is completed in $\max_i T_{i,t}$ time, when all workers send their gradient updates. We assume that the time taken by a worker $i$ to compute gradient of one mini-batch is random and independently distributed across mini-batches and workers \cite{Dutta:arxiv:18}. Specifically, we assume that $T_{i,t}$ is exponentially distributed with mean $\frac{P_i}{c_i}$, where $c_i$ denotes the total number of CPU cycles required to accomplish the computation task, and $P_i$ denotes CPU power, i.e., the computation ability represented as CPU cycles per second of worker $i$. 

The model owner negotiates with the workers about the CPU power, i.e., $P_i$. In return, each worker $i = 1,\ldots,K$ will receive the revenue $q_i P_i$ from the model owner, where $q_i$ is the price of one unit of worker $i$'s CPU power. Intuitively, the latency required to finish the learning process depends on the total usage of CPU power of all workers. Specifically, the learning latency becomes smaller as the expected value of maximum of $T_{i,t}$ reduces. As a result, the model owner aims to minimize the following cost function: 
\begin{align}
    \Delta(q_i,P_i) = V \E{\max_i T_{i,t}} + \sum_{i=1}^K q_i P_i, 
\end{align}
where $V\geq 0$ is a positive constant optimization parameter. Note that $\E{\max_i T_{i,t}}$ decreases with increasing value of $P_i$.  Let $B$ denote the maximum allowable budget of the model owner to pay for CPU power usage of cooperative workers. 

\begin{lemma}
The expected value of the time required to finish a single iteration is obtained as:
\begin{align}
 \E{\max_i T_{i,t}} = \sum_{S \subseteq \{ 1,2,\ldots,K \}} (-1)^{|S|-1}\frac{1}{\sum_{i \in S} \lambda_i},
\end{align}
where the outer sum is over all non-empty subsets $S$ of $\{ 1,2,\ldots,K \}$ and $|S|$ denotes the number of elements of $S$. In addition, $\lambda_i = \frac{P_i}{c_i}$.

\proof {We omit the proof due to lack of space, but it follows the same lines of derivation of Proposition 3.2 in \cite{Bibinger13}. }
\label{lemma:maksimum_time}
\end{lemma}

Note that as the workers obtain a revenue of $q_i P_i$ from the model owner, each model device has an energy cost incurred from the computation, which is directly dependent on the value of CPU power usage, $P_i$ as: $\kappa c_i (P_i)^2$, where $\kappa$ is a coefficient depending on the chip architecture \cite{Zhang:IoT:18}. Thus, the objective of each worker $i$ is to maximize the following utility function:
\begin{align}
    U_i(P_i,q_i) = q_i P_i -\kappa c_i (P_i)^2.\label{eq:utility_worker}
\end{align}
where $P_{\max}$ is the maximum of allowable CPU power usage of the workers. 
\section{Stackelberg Game Formulation and Equilibrium Solution}
Next, we model the interaction between the workers and the model owner as a Stackelberg game. In the lower level of the game, the workers determine their CPU power, $P_i$, as a function of price per unit, $q_i$. In the upper level, the model owner decides on the price per unit power for each worker, $q_i$. 
%Moreover, since the workers cooperatively send their gradient update to the model owner, each worker also needs to independently decide on its CPU power usage. 
As a result, the Stackelberg game can be formally defined as follows: 

1) {\bf Lower-level Subgame}: Given the fixed vector of prices of one unit of CPU power $\boldsymbol{q} = q_1, q_2, \ldots, q_K$, the lower-level subgame problem is defined as:
\begin{align}
    \max_{P_i} \ \ & U_i (P_i,q_i)= q_i P_i -\kappa c_i (P_i)^2 \\ 
    \mbox{subject to} \ \ & P_i \leq P_{\max},
\end{align}
where $P_{\max}$ is the maximum available CPU power. 

2) {\bf Upper-level Subgame}: After each worker's CPU power utilization with respect to prices, the model owner forms a upper-level subgame problem as defined as:
\begin{align}
    \min_{\boldsymbol{q}} \ \ & \Delta = V \E{\max_i T_{i,t}} + \sum_{i=1}^K q_i P_i \label{objective:higher_level}\\
    \mbox{subject to} \ \ & \sum_{i=1}^K q_i P_i \leq B \label{constraint:higher_level},
\end{align}
where $B$ is the available budget to pay the workers. 

Based on the game formulation, we consider a Stackelberg equilibrium to the solution for the model owner and the workers. Specifically, by following the backward induction, we firstly use the first-order optimality condition to obtain the optimal solution to the lower level subgame. Then, we substitute the Nash equilibrium of the lower-level subgame into the upper-level subgame and investigate the solution to the upper-level subgame. 
\vspace{-0.8cm}
\subsection{Solution to Lower-level Subgame}
To find the optimal solution for the lower-level subgame at each worker, we take the first derivative of the utility function of each worker in \eqref{eq:utility_worker} with respect to $P_i$:
\begin{align}
    \frac{\partial}{\partial P_i} U_i(P_i,q_i)= \frac{\partial}{\partial P_i} \left[ q_i P_i -\kappa c_i (P_i)^2 \right] = q_i - 2 \kappa c_i P_i, \forall i 
    \label{eq:stat-cond}
\end{align}
By equating \eqref{eq:stat-cond} to zero, we obtain the optimal CPU power as:
\begin{align}
    P_i^*(q_i) = \begin{cases} &\frac{q_i}{2\kappa c_i} \mbox { if } \frac{q_i}{2\kappa c_i} \leq P_{\max} \\ 
        & P_{\max}   \mbox { if } \frac{q_i}{2\kappa c_i} > P_{\max}       \end{cases}
        \label{eqaution:optimal_power}
\end{align}

Furthermore, it is easy to show that the utility function of each worker is strictly concave, which guarantees the existence and uniqueness of nash equilibrium. 

\subsection{Solution to Upper-level Subgame}

After obtaining the optimal CPU power of each worker as a function of price per unit CPU power, we investigate the solution to the upper-level subgame for the model owner. Due to the high non-linearity of the maximum time equation given in Lemma \ref{lemma:maksimum_time}, we cannot obtain the closed form solution for the general case. Instead, we present Lemma \ref{lemma:boundary} that can be used to develop an efficient update algorithm to reach the equilibrium point for heterogeneous case, where $c_i \neq c_j, \ \forall i,j \in \{ 1,2, \ldots, K\}$. 
\begin{lemma}
When $V$ is sufficiently large, the optimal solution is realized at the boundary, i.e., $\sum_{i=1}^K \frac{(q_i^*)^2}{2\kappa c_i} = B$, where $q_i^*$ is optimal budget allocation per unit of power for worker $i$.
\label{lemma:boundary}

 \proof {The proof is given in Appendix \ref{appendix:boundary}}.
\end{lemma}

The closed-form solution to the homogeneous case where $c_i = c_j, \ \forall i,j \in \{ 1,2, \ldots, K\}$ is given in the following theorem. 
\begin{theorem}
    When $c_i = c_j = c, \ \forall i,j \in {1,2, \ldots, K}$, the optimal solution to the upper-level subgame is $q_i^* = \sqrt{ \frac{2 B \kappa c}{K}},$ for all $i$.
    \proof {The proof is given in Appendix \ref{appendix:optimal_homo}}.
    \label{theorem:Upper_Soln}
\end{theorem}

%% file: numerical.tex
\section{Numerical Results}
%\vspace{-1cm}
\begin{figure*}[t!]
\centerline{ \subfloat[Performance with respect to the number of workers, $K$,]{\includegraphics[width=3.5in]{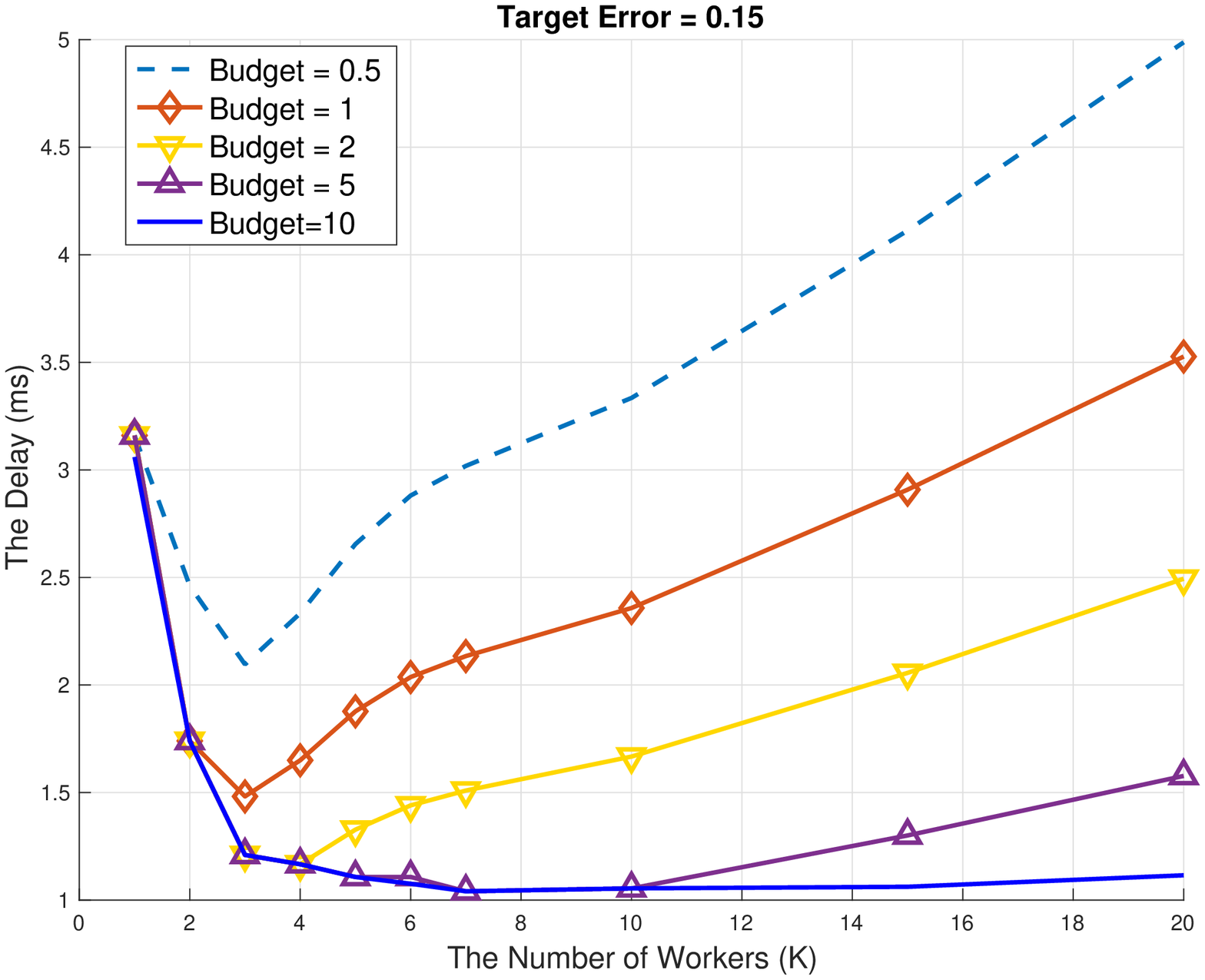}
\label{fig:K}}  
\subfloat[The budget of the model-owner vs.\ the optimal number of workers.]{\includegraphics[width=3.5in]{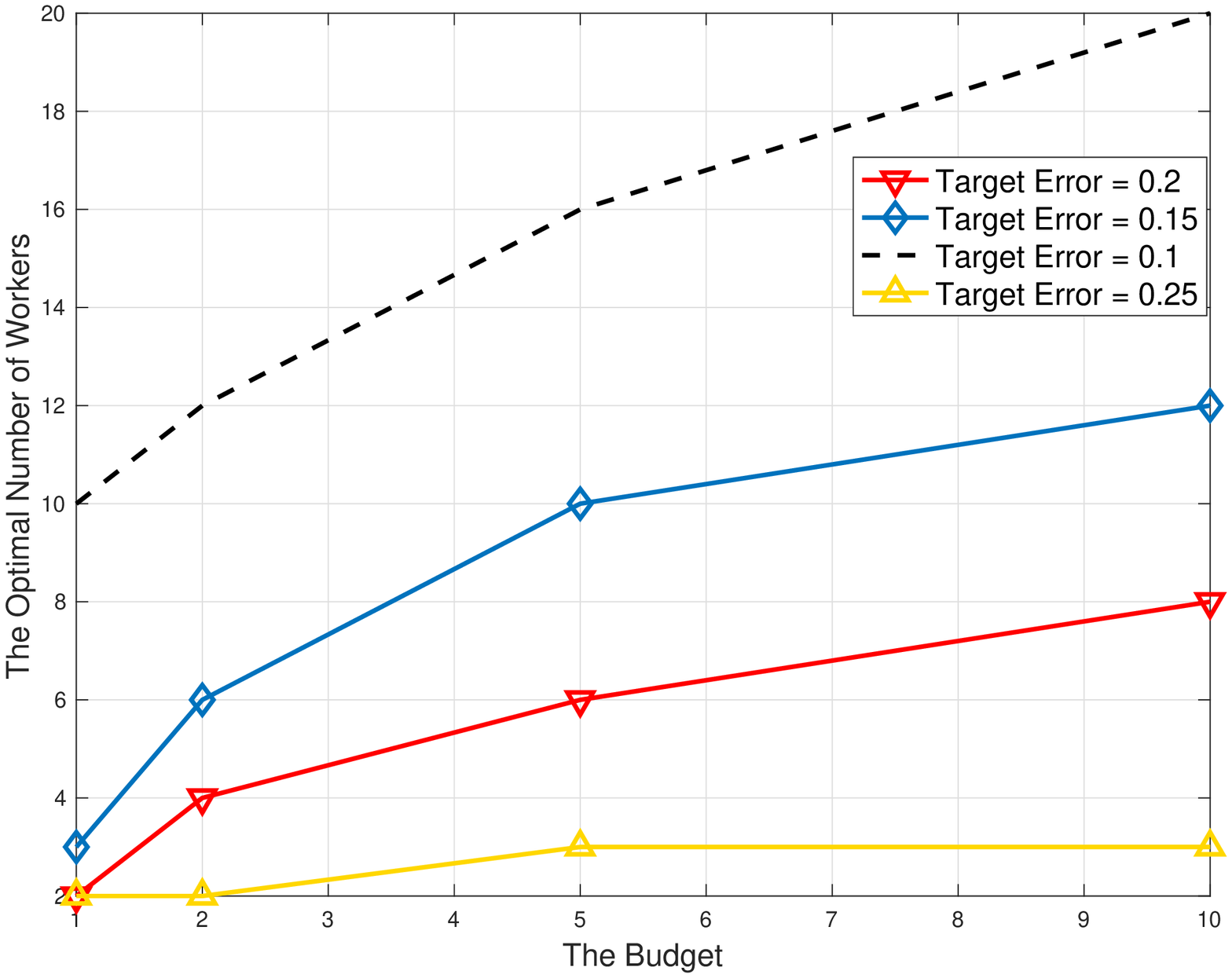}
\label{fig:B}}} \caption{Performance analysis of the Stackelberg federated learning game model.}
\end{figure*}

In the simulations, we use MNIST dataset for which we first convert the 28 x 28 images into single vectors of length 784. We use a single layer of neurons followed by soft-max cross entropy with logits loss function. Thus, effectively the parameters consist of a weight matrix $\boldsymbol{W}$ of size 784 x 10 and a bias vector $\boldsymbol{b}$ of size 1 x 10. We use a regularizer of value 0.01, and learning rate of value 0.05. For implementation we used Tensorflow with Python3. For the run-time simulations, we generate random variables from the respective distributions in python to represent the computation times. Specifically, the computation time for each worker is generated from an exponential distribution with mean $\frac{P_i}{c_i}$. Furthermore, to consider heterogeneous workers in the system, we select $c_i$ uniformly at random in the range of $[0.5 \cdot 10^3, 1.5 \cdot 10^3]$.  Due to the randomness of the selection of $c_i$ and randomness of stochastic gradient descent (SGD), we run each realization 50 times and take the average. In the simulations, we are interested in the error rate defined as ratio of the difference of the processed image and the original image with the original image. In all simulations, we defined a target error rate, and if the target error rate is realized, we stop the simulations and consider the time elapsed to reach the pre-defined error rate. 

We first investigate the effect of varying number of workers and budget on the latency in Fig. \ref{fig:K}. In all budget values, the latency initially decreases with the number of workers, since the error improves with increasing $K$, and thus, the number of workers in the training data set process leads to increase diversity and to reach the target error value, the system requires fewer iterations. However, after a certain point, the latency starts to increase. This is due to the fact that as the number of helpers increases, the positive effect of diversity of training data diminishes and the delay resulted in waiting for the update of all workers, starts to dominate. We also observe that the time required to reach the target error rate decreases, as the budget of the model owner increases. This is because an increase in the budget results in more CPU power allocated per worker reducing the time to complete each iteration. As a result, the total latency decreases. Fig. \ref{fig:K} demonstrates the trade-off between the diversity, which leads to reduction in total number of iterations, and the time elapsed to complete each iteration, both of which increases with the number of workers. Henceforth, for a given budget and target error rate, there exists an optimal number of workers that should be employed by the model owner. 

Next, we investigate the optimal number of workers minimizing the total latency for varying budget and target error rates. As depicted in Fig. \ref{fig:B}, an increase in the budget leads to an increase in the optimal number of workers, since as the budget increases, more CPU power can be purchased from more workers. Furthermore, as target error rate decreases, the optimal number of workers increases. This is because as the target error rate decreases, the number of iterations to complete the process increases, which in turn allows the diversity of training data provided by different workers to become more effective. 

%In summary, our analysis informs the model-owner about the optimal choice of $K$ to achieve a good error-latency trade-off. 

%% file: conclusion.tex
\section{Conclusion}
In this paper, we have presented a Stackelberg game model to analyze the CPU allocation strategies of multiple workers as well as the budget allocation of the model owner in a synchronous SGD run by the model owner. %We have focused on the interactions between the workers and the model owner.
Specifically, we have investigated the impact of the available budget and target error rate on CPU power utilization of workers and the convergence time of the learning process. We observe that even though higher number of workers leads to higher diversity in the learning process, there is a maximum number of workers beyond which the delay due to waiting for SGD update dominates. This result demonstrates the importance of an efficient resource allocation algorithms in a practical learning system. 

One important direction of extension of this work is to consider a dynamic game formulation that arises when the dynamic channel and worker CPU conditions are taken into account.  Additionally, the interactions between the model owner and workers depend on the learning approach implemented, e.g., AdaGrad, ADAM, etc. Although similar trade-offs exist regardless of the method implemented, it would be insightful to study the optimal number of workers depending on the method used.
%Our future work will extend to the study with varying CPU capabilities of workers. In addition, we plan to incorporate communication in the model as well. In this case, besides the learning process, power allocation in communication processes will make resource competition more challenging. In addition, the common communication environment will require the use of intelligent access control algorithms. Finally, because of the mobility, the communication channels will change dynamically, in which case dynamic control algorithms and dynamic game theory will need to be considered. 

%% file: appendix_boundary.tex
\section{Proof of Lemma \ref{lemma:boundary}}
\label{appendix:boundary}
We first substitute the optimal CPU power allocations for the workers, i.e., $P_i^*$, given as in \eqref{eqaution:optimal_power} into the cost minimization problem given in \eqref{objective:higher_level} - \eqref{constraint:higher_level}. As the constraint in \eqref{constraint:higher_level} is linear, we adopt the Lagrangian method. The Lagrangian function for the optimization problem \eqref{objective:higher_level} - \eqref{constraint:higher_level} is given as follows:
\begin{align}
    L(\boldsymbol{q},\alpha) = V \E{\max_i T_{i,t}} + \sum_{i=1}^K \frac{q_i^2}{2\kappa c_i} + \alpha \left( \sum_{i=1}^K \frac{q_i^2}{2\kappa c_i} - B \right),
\end{align}
where $L(\cdot)$ is Lagrangian function and $\alpha$ denotes Lagrangian multiplier. 

To obtain the optimal solution, we take the first derivative of Lagrangian function with respect to $q_i$.
\begin{align}
    \frac{\partial L(\boldsymbol{q},\alpha)}{\partial q_i} &= V \frac{\partial \E{\max_i T_{i,t}}}{\partial q_i} + \frac{q_i}{\kappa c_i} + \alpha \frac{q_i}{\kappa c_i}, \ \forall i \nonumber \\
     &= V\frac{\partial \E{\max_i T_{i,t}}}{\partial \lambda_i}\frac{1}{2 \kappa c_i^2} + 2 \lambda_i c_i + 2\alpha \lambda_i c_i, \ \forall i.
    \label{eq:alpha_3}
\end{align}
In \eqref{eq:alpha_3}, we use the relation between $\lambda_i$ and $q_i$, i.e. $\lambda_i = \frac{q_i}{2\kappa c_i^2}$. Then, we equate the first derivative given in  \eqref{eq:alpha_3} to zero to derive the value of Lagrange multiplier, $\alpha$, at the optimal point as:
\begin{align}
    \alpha = - V\frac{\partial \E{\max_i T_{i,t}}}{\partial \lambda_i} \frac{1}{4\kappa c_i \lambda_i} -1, \ \forall i.
    \label{eq:alpha_4}
\end{align}

Similarly, from the second and third  Karush-Kuhn-Tucker (KKT) conditions we have $\alpha \left( \sum_{i=1}^K \frac{q_i^2}{2\kappa c_i} - B \right) = 0$ and $\alpha \geq 0 $. Thus, the first term in \eqref{eq:alpha_4} should be positive. Intuitively, as the exponential parameter, i.e., the inverse of mean completion time, $\lambda_i$, increases, the maximum value of completion times should decrease. As a result, for a sufficiently large $V$, we can guarantee that $\alpha$ is positive. Thus, from complementary slackness condition of KKT, the solution exists at the boundary.

%% file: appendix_Theorem.tex
\section{Proof of Theorem \ref{theorem:Upper_Soln}}
\label{appendix:optimal_homo} 

Using the result given in \eqref{eq:alpha_3} and the fact that $c_i = c_j = c, \ \forall i,j$, we obtain the following relation: 
\begin{align}
    \frac{\partial \E{\max_i T_{i,t}}}{\partial \lambda_i} \frac{1}{\lambda_i} &= \frac{\partial \E{\max_i T_{i,t}}}{\partial \lambda_j} \frac{1}{\lambda_j}, \forall i,j \nonumber \\
    \sum_{S \subseteq \{ 1,2,\ldots,n \} 
    \text{ and } i \in S} (-1)^{|S|}\frac{1}{\left(\sum_{k \in S} \lambda_k\right)^2\lambda_i} &= \sum_{S \subseteq \{ 1,2,\ldots,n \}  \mbox{ and } j \in S} (-1)^{|S|}\frac{1}{\left(\sum_{k \in S} \lambda_k\right)^2\lambda_j} \nonumber \\ 
    \sum_{S \subseteq \{ 1,2,\ldots,n \}
    \text{ and } i \in S, j \notin S} (-1)^{|S|}\frac{1}{\left(\sum_{k \in S} \lambda_k\right)^2\lambda_i} &= \sum_{S \subseteq \{ 1,2,\ldots,n \}  \mbox{ and } j \in S, i\notin S} (-1)^{|S|}\frac{1}{\left(\sum_{k \in S} \lambda_k\right)^2\lambda_j}
    \label{eq:equality_3}
\end{align}

A trivial solution to \eqref{eq:equality_3} is achieved when $\lambda_i = \lambda_j = \lambda, \ \forall i,j \in{1,2, \ldots, K}$. This will also result in equal value of $q_i$'s. Combining this result and the result given in Lemma \ref{lemma:boundary}, we obtain:
\begin{align*}
\frac{q_i^2}{2 \kappa c_i } = \frac{B}{K} \ \ 
\Rightarrow \ \ q_i^* = \sqrt{\frac{2B \kappa c}{K}}.
\end{align*}

%% file: main.bbl
\begin{thebibliography}{10}
\providecommand{\url}[1]{#1}
\csname url@samestyle\endcsname
\providecommand{\newblock}{\relax}
\providecommand{\bibinfo}[2]{#2}
\providecommand{\BIBentrySTDinterwordspacing}{\spaceskip=0pt\relax}
\providecommand{\BIBentryALTinterwordstretchfactor}{4}
\providecommand{\BIBentryALTinterwordspacing}{\spaceskip=\fontdimen2\font plus
\BIBentryALTinterwordstretchfactor\fontdimen3\font minus
  \fontdimen4\font\relax}
\providecommand{\BIBforeignlanguage}[2]{{%
\expandafter\ifx\csname l@#1\endcsname\relax
\typeout{** WARNING: IEEEtran.bst: No hyphenation pattern has been}%
\typeout{** loaded for the language `#1'. Using the pattern for}%
\typeout{** the default language instead.}%
\else
\language=\csname l@#1\endcsname
\fi
#2}}
\providecommand{\BIBdecl}{\relax}
\BIBdecl

\bibitem{McMahan16}
H.~B. McMahan, E.~Moore, D.~Ramage, S.~Hampson, and B.~A. Arcas,
  ``{Communication-efficient learning of deep networks from decentralized
  data},'' in \emph{International Conference on Artificial Intelligence and
  Statistics (AISTATS)}, Fort Lauderdale, FL, USA, Apr. 2017.

\bibitem{amazon}
A.~G.~D. R.~Tandon, Q.~Lei and N.~Karampatziakis, ``Gradient coding: avoiding
  stragglers in distributed learning,'' in \emph{Proc. Int. Conf. on Machine
  Learning}, Sydney, Australia, Feb. 2017, pp. 3368--3376.

\bibitem{Lee:TIT:18}
K.~Lee, M.~Lam, R.~Pedarsani, D.~Papailiopoulos, and K.~Ramchandran, ``Speeding
  up distributed machine learning using codes,'' \emph{IEEE Transactions on
  Information Theory}, vol.~64, no.~3, pp. 1514--1529, 2018.

\bibitem{Karakus17}
C.~Karakus, Y.~Sun, S.~Diggavi, and W.~Yin, ``{Straggler mitigation in
  distributed optimization through data encoding},'' in \emph{Advances in
  Neural Information Processing Systems 30 (NIPS)}, Long Beach, NY, USA, Dec.
  2017, pp. 5440--5448.

\bibitem{Harlap16}
A.~Harlap, H.~Cui, W.~Dai, J.~Wei, G.~R. Ganger, P.~B. Gibbons, G.~A. Gibson,
  and E.~P. Xing, ``{Addressing the straggler problem for iterative convergent
  parallel ml},'' in \emph{ACM Symposium on Cloud Computing (SoCC)}, Santa
  Clara, CA, USA, Oct. 2016, pp. 98--111.

\bibitem{Sun:arxiv:19}
\BIBentryALTinterwordspacing
Y.~Sun, J.~Zhao, S.~Zhou, and D.~G{\"{u}}nd{\"{u}}z, ``Heterogeneous
  computation across heterogeneous workers,'' \emph{CoRR}, vol. abs/1904.07490,
  2019. [Online]. Available: \url{http://arxiv.org/abs/1904.07490}
\BIBentrySTDinterwordspacing

\bibitem{Mohammad:arxiv:18}
\BIBentryALTinterwordspacing
U.~Mohammad and S.~Sorour, ``Adaptive task allocation for mobile edge
  learning,'' \emph{CoRR}, vol. abs/1811.03748, 2018. [Online]. Available:
  \url{http://arxiv.org/abs/1811.03748}
\BIBentrySTDinterwordspacing

\bibitem{Feng:arxiv:18}
\BIBentryALTinterwordspacing
S.~Feng, D.~Niyato, P.~Wang, D.~I. Kim, and Y.~Liang, ``Joint service pricing
  and cooperative relay communication for federated learning,'' \emph{CoRR},
  vol. abs/1811.12082, 2018. [Online]. Available:
  \url{http://arxiv.org/abs/1811.12082}
\BIBentrySTDinterwordspacing

\bibitem{Dutta:arxiv:18}
\BIBentryALTinterwordspacing
S.~Dutta, G.~Joshi, S.~Ghosh, P.~Dube, and P.~Nagpurkar, ``Slow and stale
  gradients can win the race: Error-runtime trade-offs in distributed sgd,''
  \emph{CoRR}, vol. abs/1803.01113, 2018. [Online]. Available:
  \url{https://arxiv.org/abs/1803.01113}
\BIBentrySTDinterwordspacing

\bibitem{Bibinger13}
\BIBentryALTinterwordspacing
M.~Bibinger, ``Notes on the sum and maximum of independent exponentially
  distributed random variables with different scaleparameters,'' \emph{CoRR},
  vol. abs/1307.3945, 2013. [Online]. Available:
  \url{http://arxiv.org/abs/1307.3945}
\BIBentrySTDinterwordspacing

\bibitem{Zhang:IoT:18}
J.~{Zhang}, X.~{Hu}, Z.~{Ning}, E.~C.~. {Ngai}, L.~{Zhou}, J.~{Wei},
  J.~{Cheng}, and B.~{Hu}, ``Energy-latency tradeoff for energy-aware
  offloading in mobile edge computing networks,'' \emph{IEEE Internet of Things
  Journal}, vol.~5, no.~4, pp. 2633--2645, Aug 2018.

\end{thebibliography}
